\documentstyle[12pt]{article}

\begin{document}

\font\ninerm = cmr9



\def\footnoterule{\kern-3pt \hrule width \hsize \kern2.5pt}

\pagestyle{empty}


\vskip 0.5 cm



\begin{center}
{\large\bf Comparison of relativity theories with\\
observer-independent  scales of both velocity and length/mass}
\end{center}

\vskip 1.5 cm

\begin{center}
{\bf Giovanni AMELINO-CAMELIA}$^{a,b}$, {\bf Dario BENEDETTI}$^{a}$,
{\bf Francesco D'ANDREA}$^{a}$ and {\bf Andrea PROCACCINI}$^{a}$\\
\end{center}

\begin{center}
$^{a}${\it Dipart.~Fisica,
Univ.~Roma ``La Sapienza'',
P.le Moro 2, 00185 Roma, Italy}\\
$^{b}${\it Perimeter Institute for Theoretical Physics,
Waterloo, Canada N2J 2W9}
\end{center}

\vspace{1cm}

\begin{center}

{\bf ABSTRACT}

\end{center}

{\leftskip=0.6in \rightskip=0.6in
We consider the two most studied proposals of
relativity theories with observer-independent scales of both velocity
and length/mass: the one discussed by
Amelino-Camelia as illustrative example for
the original proposal (gr-qc/0012051)
of theories with two relativistic invariants,
and an
alternative more recently proposed by Magueijo and Smolin (hep-th/0112090).
We show that these two relativistic theories are much more closely
connected than it would appear on the basis of a naive analysis
of their original formulations. In particular, in spite of adopting a rather
different formal description of the deformed boost generators,
they end up assigning the same dependence of momentum on rapidity,
which can be described as the core feature of these relativistic theories.
We show that this observation can be used to clarify
the concepts of particle mass, particle velocity,
and energy-momentum-conservation rules in these theories
with two relativistic invariants.
}

\newpage


\baselineskip 16pt plus .5pt minus .5pt

\pagenumbering{arabic}

\pagestyle{plain}


\section{Introduction}
One of us put forward in Refs.~\cite{amel-dsr1,amel-dsr2}
the proposal of special-relativistic theories
(theories of the transformation rules that connect the
observations of different {\underline{inertial}} observers)
with {\underline{two}} observer-independent scales.
Galilei's relativity principle peacefully coexists with
the absence of observer-independent scales, as shown by
the structure of the Galilei-Newton transformation rules.
Einstein's Special Relativity relies on the ordinary Lorentz transformations,
which host one observer-independent scale, the velocity scale $c$.
In Refs.~\cite{amel-dsr1,amel-dsr2}
it was argued that there should also be some examples of
special-relativistic theories with two (or more)
observer-independent scales,
which could be called~\cite{amel-dsr1} ``Doubly
Special Relativity" or ``DSR",
and a first example, which we shall call DSR1 in the following,
was analyzed in detail, including a careful formulation of the postulates,
an explicit derivation of the {\underline{finite}}
deformed Lorentz transformations
and a study of the kinematical
conditions for particle production in collision processes.
Most of the quantitative results reported in Refs.~\cite{amel-dsr1,amel-dsr2}
were obtained in leading order (all orders in $c \sim 3 {\cdot} 10^8 m/s$
but only leading order
in the second observer-independent scale, tentatively identified
with the Planck length $L_p \sim 1.6 {\cdot} 10^{-35}m$
or the corresponding Planck mass $E_p \equiv 1/L_p$)
and were later generalized to all orders in
Refs.~\cite{amel-jurekdsr,amel-starkappa,amel-gacJR,polon2pap}.
While progress was coming quickly in the analysis of the first DSR
example, for some time no other DSR example was identified in the
literature. This changed when Magueijo and Smolin
proposed another DSR theory in Ref.~\cite{leedsr}, which we shall
call DSR2 in the following.
More recently, other possible realization of the DSR idea
have been considered (see, {\it e.g.}, Refs.~\cite{jurekDSRexamples,dsrgzk}),
but DSR1 and DSR2 remain the focus of the majority
of what is becoming a rather large DSR literature (see, {\it e.g.},
Refs.~\cite{judesvisser,polonDSR2nogood,dsrNATURE,chakdsr,ukDSR2,belgiumDSR2})

DSR1 and DSR2 were originally formulated using somewhat
different conventions and notation, and are being considered
in the literature as significantly different~\cite{judesvisser}
realizations of the DSR idea; most studies focusing
one or the
other~\cite{polonDSR2nogood,dsrNATURE,chakdsr,ukDSR2,belgiumDSR2}.
But in the literature one can still not find a robust comparative
study of these two relativistic theories.
Our analysis will here focus on establishing the common features and the main
differences between these two DSR proposals.
We will argue that a DSR theory is primarily characterized
by the dependence of (space) momentum on rapidity.
The only other necessary ingredient is the $E(p)$ (energy/momentum)
dispersion relation (once the dependence of $p_i$ on rapidity is
fixed, the dispersion relation also fixes the dependence of
energy on rapidity).
And we find that remarkably, in spite of adopting a rather
different formal description of the deformed boost generators,
DSR1 and DSR2 predict the same
dependence of momentum on rapidity.
We therefore expose the fact that the difference between these
two relativistic theories amounts to a simple
redefinition of energy, a single function of a single variable,
whereas it was so far assumed
that some nontrivial nonlinear transformation $P_\mu^{DSR2}(P_\mu^{DSR1})$
(four functions, each depending on four variables)
should be involved in the connection between DSR1 and DSR2.

In light of the simplicity of the relation between DSR1 and DSR2,
sharing the same core relativistic features,
we find appropriate to reanalize some of the perspectives
on these two theories which have been presented in the literature.
In particular,
some studies considering DSR2 have adopted a description of
particle mass and of particle velocity
which is different from the one adopted in most studies
of DSR1. Since we are finding that DSR1 and DSR2 are connected in
an elementary way, it seems illogical that such differences in
the conceptual analysis would be necessary.
Indeed we will find that these claimed differences are due
to misinterpretation of some differences which are purely in
the realm of the choice of notation, and therefore
cannot have physical consequences.

We also provide a careful analysis of kinematical
conditions for particle production in collision processes,
which are a key component of DSR theories.
These kinematical conditions were already carefully analyzed
within the DSR1 proposal in
Refs.~\cite{amel-dsr1,amel-dsr2,amel-starkappa,amel-gacJR,polon2pap},
whereas in the paper~\cite{leedsr} announcing the DSR2 proposal
there was no discussion of these kinematical conditions.

Importantly we find that, although the difference between DSR1 and DSR2
is of marginal conceptual significance,
there are some contexts in which from a quantitative
perspective the two theories lead to significantly different predictions.
Research conducted over the last few
years~\cite{domokos,amel-grbgac,glast,amel-kifu,gacQM100,amel-gactp2}
has led to the conclusion that for theories, like DSR1 and DSR2,
predicting a Planck-scale deformation
of the dispersion relation $E(p)$,
the only two contexts in which
the new effects could be observably large are: (i) astrophysical
studies of signal dispersion, which are sensitive to the relativistic
relation between velocity and
momentum~\cite{domokos,amel-grbgac,glast,gacQM100}, and (ii)
analyses of cosmic-ray data, which are sensitive to the structure
of the relativistic laws of energy-momentum
conservation~\cite{amel-kifu,gacQM100,amel-gactp2}.
From the results here reported we conclude
that DSR2 (just as previously established for DSR1~\cite{amel-dsr1,amel-dsr2})
only has negligibly small implications for cosmic-ray physics.
We also find that the DSR2 relation between velocity and momentum
of a particle does not lead to observably large new effects, while the
corresponding DSR1 prediction is testable and will be tested~\cite{glast}
exploiting the remarkable sensitivity~\cite{amel-grbgac} of
the GLAST space telescope.

Since the comparative study we are reporting is
articulated over several points, for clarity we reserve the next Section
to a description of the key characteristics of DSR1,
while Section~3 discusses the corresponding characteristics
of DSR2 and compares the two relativistic theories.
Section~4 is an aside on the nonlinearities in
the DSR framework.
The closing Section~5
summarizes our key results and
emphasizes some key open issues.

\section{Main features of DSR1}

\noindent
{\bf (DSR1.a): general structure.}
As mentioned, work on DSR1 started already in the
original papers~\cite{amel-dsr1,amel-dsr2}
that proposed the idea of theories with
both a velocity scale and a length/momentum scale
as relativistic invariants.
The analysis preliminary analysis of DSR1
presented in Refs.~\cite{amel-dsr1,amel-dsr2} intended to show
that Galilei's Relativity
Principle can coexist with postulates introducing
two observer-independent scales.

The guiding intuition came in part from research in quantum gravity
where it is often assumed that the Planck length $L_p$ has a
fundamental role in the short-distance structure of space-time.
Such a structural role for the Planck length can easily
come into conflict with one of the cornerstones
of Einstein's Special Relativity:
FitzGerald-Lorentz length contraction.
According to FitzGerald-Lorentz length contraction,
different inertial observers would attribute different
values to the same physical length.
If the Planck length only has the role we presently attribute
to it, which is basically the role of a coupling constant
(an appropriately rescaled version of the gravitational coupling),
no problem arises for FitzGerald-Lorentz contraction,
but if we try to promote $L_p$ to the status of an intrinsic
characteristic of space-time structure
it is natural to find conflicts with FitzGerald-Lorentz contraction.
For example, it is very hard (perhaps even impossible)
to construct discretized versions or non-commutative versions
of Minkowski space-time which enjoy ordinary
Lorentz symmetry.\footnote{Pedagogical illustrative examples of
this observation have been discussed, {\it e.g.},
in Ref.~\cite{amel-hooftlorentz} for the case of discretization
and in
Refs.~\cite{amel-kpoinfirst,amel-rueggnew,amel-majrue,amel-kpoinap}
for the case of non-commutativity.}
Therefore, unless the Relativity postulates are modified,
it appears impossible to attribute to the Planck length
a truly fundamental (observer-independent) intrinsic
role in the microscopic structure of space-time.

It is clearly not necessary to introduce such a modification
of the Relativity postulates, since we do not (yet?) have
any conclusive experimental evidence that require us to attribute to $L_p$
an observer-independent role in the microscopic structure of space-time,
but it is of course legitimate~\cite{amel-dsr1,amel-dsr2}
to explore this possibility.

Actually, there is some tentative encouragement from experiments
for the idea that $L_p$ is also a feature of kinematics,
rather than simply a coupling constant.
In fact, over the last few years there have been attempts
to interpret puzzling observations~\cite{uhecrdata}
of ultra-high-energy cosmic rays
as a manifestation of new rules of kinematics for particle
production in collision processes, and that these
new rules might involve a new kinematical length scale.
The relevant process is
photopion production by high-energy protons colliding
with soft photons.
These processes are important because
at sufficiently high energies, above
a threshold energy $E_{th}$,
pion-producing interactions between the high-energy
cosmic-ray proton and one of the soft photons in the
CMBR environment
become kinematically allowed.
It was shown (see, {\it e.g.},
Refs.~\cite{amel-kifu,amel-aus,amel-gactp2,gacQM100})
that certain puzzling aspects of cosmic-ray
observations could be explained if the conventional
special-relativistic
estimate of the
photopion production threshold energy was modified
at order $L_p^n E_{th}^{1+n}$, with $n \le 1$.

Another class of observations relevant for Lorentz invariance
which is improving very rapidly are the ones pertaining to
a possible wavelength/energy dependence of the speed of
photons~\cite{amel-grbgac,amel-billetal}.
With experiments such as AMS~\cite{ams} and GLAST~\cite{glast}
the expected sensitivity should allow to investigate the possibility
of corrections of order $L_p E$ to the speed-of-light
law, $v_\gamma \simeq c (1 {\pm} L_p E)$.

Since the objective of the present study is
the one of comparing the two most studied DSR proposals,
rather than providing any encouragement (or discouragement)
for the DSR idea, we refer the
interested reader to Refs.~\cite{amel-dsr1,amel-dsr2}
for a more detailed description of the motivation
and the logical structure of DSR theories.
It will become evident in the following that DSR1 proved to
be a good choice as example of relativistic theory with
two invariants to be used in illustrating the proposal
put forward in Refs.~\cite{amel-dsr1,amel-dsr2}.
In fact, DSR1 attributes to the Planck length the role
of inverse of the maximum value of momentum that can be
held by fundamental particles, and in this sense may be
an appealing possibility for work in quantum gravity
in which some sort of quantization of spacetime at the
planck scale is introduced.
Moreover, DSR1 predicts
corrections of order $L_p E$ to the speed-of-light
law and is therefore appealing from a phenomenological perspective,
since, as just mentioned, this effect can be tested in the not-so-distant
future.

\noindent
{\bf (DSR1.b): generators of deformed Lorentz transformations.}
As argued in Refs.~\cite{amel-dsr1,amel-dsr2}
the introduction of the second (length/momentum) relativistic invariant
does not naturally invite us to revise space rotations, but it
appears inevitable that such relativistic theories
should involve deformed boost generators.
The DSR1 relativistic theory involves
the following differential representation of the boost
generators (without loss of
generality we choose to focus on the boost that acts
along the $z$ axis):
\begin{equation}
N_z=  p_z
\frac{\partial}{\partial E}
+  \left(\frac{{\tilde L}_p}{2} {\vec p}^2
+ \frac{1-e^{-2 {\tilde L}_p E}}{2 {\tilde L}_p}\right)
\frac{\partial}{\partial p_z}
- {\tilde L}_p p_z \left(p_j\frac{\partial}{\partial p_j} \right )
~.
\label{generatDSR1exact}
\end{equation}
${\tilde L}_p$ is here assumed to be of the order of the Planck
length (but not necessarily given exactly by the Planck length).

We also note here, because of its usefulness for the study
of possible experimental tests of DSR1 (which would of course
be only sensitive to leading-order effects),
the formula for boost generators approximated
in leading order in the deformation scale
\begin{equation}
N_z \simeq p_z \frac{\partial}{\partial E}
+ \left(E + \frac{{\tilde L}_p}{2} {\vec p}^2
- {\tilde L}_p E^2 \right)
\frac{\partial}{\partial p_z}
- {\tilde L}_p p_z \left(p_j\frac{\partial}{\partial p_j} \right )
~.
\label{generatDSR1approx}
\end{equation}

\noindent
{\bf (DSR1.c): deformed Lorentz transformation rules.}
The DSR1 Lorentz generators of course give a direct description
of infinitesimal Lorentz transformations.
In order to obtain finite Lorentz transformations (the ones
that truly describe Lorentz symmetry in physics)
one needs to exponentiate these generators, thereby constructing
candidate elements of a group of (deformed) Lorentz transformations.
The generators (\ref{generatDSR1exact}) had already surfaced
in the context of math-oriented studes
of $\kappa$-Poincar\`{e} Hopf algebras.
It was well known that
for Lorentz-sector generators
of generic $\kappa$-Poincar\`{e} Hopf algebras
the exponentiation procedure does not actually lead to a group
(one only obtains a quasigroup in the sense of
Batalin~\cite{amel-rueggnew,batalin}).
For the construction of DSR1 it was therefore a key technical point,
from the math perspective, the observation~\cite{amel-dsr1}
that the generators of
the specific Lorentz sector of $\kappa$-Poincar\`{e} Hopf algebra
described in (\ref{generatDSR1exact})
do lead to group structure\footnote{The emerging group is just the
ordinary Lorentz group (same composition law) but nonlinearly
realized. This is easily verified~\cite{amel-dsr1}
using the Baker-Campbell-Hausdorff
formula (applied to group elements constructed as exponentials
of the generators)
and exploiting the fact that the deformed Lorentz generators
still satisfy the ordinary Lorentz algebra.}
upon exponentiation.
The DSR1 generators are therefore consistent with a genuine
group of deformed Lorentz symmetries, whereas generic boost generators
within $\kappa$-Poincar\`{e} Hopf algebras are not.

For the purposes of the
later comparison between DSR1 and DSR2 it is convenient to
revisit here briefly how from the DSR1 generators one
obtains the DSR1 transformation laws between inertial observers.
This was discussed in detail in
Refs.~\cite{amel-dsr1,amel-dsr2,amel-gacJR}.
The mentioned exponentiation of generators actually means
that energy and momentum satisfy some differential
equations as function of rapidity. From (\ref{generatDSR1exact})
one straightforwardly obtains
(for simplicity, in the case in which
the components of momentum orthogonal to the boost
vanish)
\begin{equation}
\left\{
\begin{array}{l}
\frac{\partial p(\xi)}{\partial
\xi}=p'(\xi)=-\frac{{\tilde L}_p}{2}p^2(\xi)+\frac{1-e^{-2 {\tilde L}_p
E(\xi)}}{2 {\tilde L}_p}\\
\\
\frac{\partial E(\xi)}{\partial \xi}=E'(\xi)=p(\xi)\\
\end{array}
\right.
\label{sist1}
\end{equation}
Here the rapidity $\xi$ is as usual intended as
the coefficient of the boost generator
in the the exponential that describes a finite group transformation.

Eq.~(\ref{sist1}) is a system of two first-order differential
equations in $p(\xi)$ and $E(\xi)$. They can be combined
to obtain a single (second-order) differential equation for $p(\xi)$
\begin{equation}
p''+{\tilde L}_p^2 p^3 + 3  {\tilde L}_p p p'-p = 0~.
\label{p}
\end{equation}
This equation can be straightforwardly integrated~\cite{amel-gacJR}.
For our purposes it is sufficient
to note here the result that describes these finite transformations
in a simple illustrative case\footnote{More general discussions of
the transformation rules can be found
in Refs.~\cite{amel-dsr1,amel-dsr2,amel-gacJR}.}:
we consider a finite Lorentz transformation
which is purely a boost in the $z$ direction and acts on a four-momentum
with components $(0,0,p_{z,0},E_0)$.
One finds the solution~\cite{amel-gacJR}
\begin{equation}
p_z(\xi) = p_{z,0} {\cosh({\xi})+A_0\sinh({\xi}) \over 1
- {\tilde L}_p p_{z,0} (A_0-A_0\cosh({\xi})-\sinh({\xi}))}  ~,
\label{pxiDSR1}
\end{equation}
where $A_0$ codifies the information about
the initial conditions
($A_0$ is $\xi$-independent), as described in Ref.~\cite{amel-gacJR}.

\noindent
{\bf (DSR1.d): dispersion relation.}
A key characteristic of DSR1, both conceptually and phenomenologically,
is the deformed dispersion relation.
Combining the deformed boosts of (\ref{generatDSR1exact})
and undeformed space-rotation generators one finds that the
 dispersion relation (the relation between energy and momentum
 which is kept invariant under the rotation/boost transformations)
 is given by
\begin{equation}
m^2 = {\tilde L}_p^{-2}\left(e^{{\tilde L}_p E}
+e^{-{\tilde L}_p E}-2\right)-\vec{p}^2 e^{{\tilde L}_p E}
~.
\label{dispexact}
\end{equation}
Especially for phenomenological applications (experimental searches)
the key point is the leading correction to the ordinary special-relativistic
dispersion relation that follows from (\ref{dispexact})
\begin{equation}
m^2 \simeq E^2 -\vec{p}^2 - {\tilde L}_p \vec{p}^2 E
~.
\label{dispapprox}
\end{equation}

Of course, using the dispersion relation, (\ref{dispexact}),
one can obtain the formula that gives the dependence
of energy on rapidity
from the formula, (\ref{pxiDSR1}), that gives the dependence
of (space) momentum on rapidity.
Therefore a compact way to codify the content of a DSR theory
is provided by a pair of equations: a formula, of type
(\ref{pxiDSR1}),  for the dependence
of momentum on rapidity and a formula, of type
(\ref{dispexact}),  that expresses energy
as a function of momentum.

\noindent
{\bf (DSR1.e): mass.}
In (\ref{pxiDSR1}) we have adopted the notation ``$m^2$" in the
dispersion relation by simple analogy with the
corresponding special-relativistic
equation.
In Einstein's special relativity the quadratic mass casimir $m^2$
coincides with the square of the rest energy $M$
and we commonly refer to both concepts as if they
were equivalent.
In the new type of relativistic theories here of interest
this correspondence is lost. In particular, in DSR1,
the rest energy, $M$ (which we shall also sometimes denote by $E_{rest}$),
is related to the $m$
of (\ref{dispexact}) by the relation
\begin{equation}
m = {\tilde L}_p^{-1} \left(e^{{\tilde L}_p M/2}
- e^{-{\tilde L}_p M/2}\right)
~.
\label{dispexactmass}
\end{equation}
Note that $M$ differs from $m$ only at order ${\tilde L}_p^2$

\noindent
{\bf (DSR1.f): maximum momentum.}
It is easy to verify from the formulas provided above
that for positive ${\tilde L}_p$
the DSR1 transformation rules
predict~\cite{amel-dsr1,amel-dsr2,amel-gacJR,polon2pap} that in
the infinite-rapidity limit the momentum
of an on-shell particle saturates to
a maximum value  $p_{max}={\tilde L}_p^{-1}$,
while energy diverges
(instead both diverge in the infinite-rapidity limit
in special relativity).

The existence of a maximum momentum in DSR1 may
suggest~\cite{amel-dsr1,amel-dsr2,polon2pap} the interpretation
of ${\tilde L}_p$ as the observer-independent minimum
value of wavelength.

\noindent
{\bf (DSR1.g): velocity.}
The law that relates the velocity of particles to their (mass and)
energy is a key aspect of a relativistic theory. Velocity is naturally
introduced in the spacetime sector, while the DSR1 postulates (just like
the DSR2 postulates described later) were formulated in energy-momentum space.
In Refs.~\cite{amel-dsr1,amel-dsr2} it was observed that
the relation $v_{particle} = {dE / dp}$ holds both in Galilei
relativity and in Einstein's special relativity, and it was then argued
that it is natural to assume the validity of this relation
also in DSR theories.
This basically amounts to the expectation that also DSR theories should
allow an Hamiltonian formulation (${\dot {x}}=dH/dp$).
It also represents a plan for the construction of the spacetime sector
of a DSR theory, a spacetime in which $v_{particle} = {dE / dp}$ would
hold~\footnote{As discussed later in the paper, it appears that
the $\kappa$-Minkowski noncommutative spacetime would provide a spacetime
realization of DSR1, and the structure of that spacetime
is indeed consistent~\cite{gacmaj} with $v_{particle} = {dE / dp}$.}.

Particularly noteworthy is the DSR1
deformed speed-of-light law\footnote{The observation
that photons of infinite energy
would have infinite velocity in DSR1
was first reported in Ref.~\cite{amel-jurekdsr}.}
\begin{equation}
v_\gamma = {dE \over dp} = e^{{\tilde L}_p E}
\simeq c \, (1 + {\tilde L}_p c^{-1} E)
~.
\label{amel-eqvelocity}
\end{equation}
This law is fully consistent with the wavelength independence
of the speed of light found in our low-energy observations
(observations at energy scales that are much smaller than $1/{\tilde L}_p$)
but would lead (see later) to observably large effects in forthcoming
experiments.


\noindent
{\bf (DSR1.h): magnitude of the deformation.}
A key aspect of a DSR theory is the magnitude of the new effects it predicts
at high energies (in the low-energy regime all realistic DSR theories
must reproduce ordinary special relativity, since Einstein's theory
is verified to very good accuracy in low-energy experiments).
Of course, for applications in phenomenology (experimental searches)
it is sufficient to establish the magnitude of the leading-order
corrections that the DSR theory predicts with respect to ordinary
special relativity. DSR1 is based on a deformed dispersion relation
which involves a correction of order ${\tilde L}_p E^3$
to the ordinary $E^2 = p^2 + m^2$ dispersion relation.
Therefore the
dimensionless quantity that characterizes the strength of
the deformation of the dispersion relation
is ${\tilde L}_p E$.
It is easy to verify that the same statement applies
to all aspects of DSR1: all the modifications of ordinary
special-relativistic results that are predicted
by DSR1 have characteristic strength ${\tilde L}_p E$.

\noindent
{\bf (DSR1.i): two-particle sector and conservation laws.}
As emphasized in Refs.~\cite{amel-dsr1,amel-dsr2,polon2pap},
in DSR theories the step from the one-particle sector
to multiparticle sectors is usually rather nontrivial.
Ordinary special relativity is a linear theory and therefore
no such complications are encountered; for example,
one can meaningfully attribute a total momentum  $\vec{p}_a + \vec{p}_b$
to a system composed of two particles, one with momentum $\vec{p}_a$
and the other with momentum $\vec{p}_b$.
The concept of total momentum is instead highly nontrivial
in DSR theories, as illustrated by the considerations
on total momentum in DSR1 reported in Ref.~\cite{polon2pap}.

For phenomenological applications a key aspect of multiparticle
systems in DSR theories are the kinematical conditions
(``energy-momentum conservation")
for particle production in collision processes.
As mentioned, certain cosmic-ray observations
might invite us to consider deformed
conservation laws.
As observed in Refs.~\cite{amel-dsr1,amel-dsr2,polon2pap},
DSR theories will necessarily involve deformed rules
for energy-momentum conservation; for example,
in a process $a + b \rightarrow c +d$ it would be inconsistent
to enforce the conditions
$E_a +E_b = E_c +E_d$, $\vec{p}_a +\vec{p}_b = \vec{p}_c +\vec{p}_d$
since these conditions would not provide an observer-independent
law. In ordinary special relativity
the conditions
$E_a +E_b = E_c +E_d$, $\vec{p}_a +\vec{p}_b = \vec{p}_c +\vec{p}_d$
can be derived from the (linear) structure of Lorentz transformations,
and they do provide an observer-independent
kinematical law for collision processes (they
are either satisfied for all inertial observers or not satisfied
for all inertial observers).
The deformed (nonlinear) transformation laws of DSR theories
would clearly not be consistent with
the conditions
$E_a +E_b = E_c +E_d$, $\vec{p}_a +\vec{p}_b = \vec{p}_c +\vec{p}_d$.

The form of the new energy-momentum-conservation-like laws
will of course depend on the structure of the specific DSR theory
(they must reflect the structure of  the transformation laws).
Much work has been done~\cite{amel-dsr1,amel-dsr2,polon2pap}
on these laws\footnote{In general
in any formalism in which some relevant nonlinearities
are introduced by a conjugation
one can derive a set of conserved quantities
by a systematic procedure~\cite{chakdsr}.}.
For the purpose of the present paper it is sufficient
to note here one example of these conservation laws
which is consistent~\cite{amel-dsr1} in leading order
in  ${\tilde L}_p$:
\begin{equation}
E_a + E_b - {\tilde L}_p c \vec{p}_a {\cdot} \vec{p}_b
- E_c -E_d + {\tilde L}_p c \vec{p}_c {\cdot} \vec{p}_d \simeq 0
~,
\label{amel-conservnewe}
\end{equation}
\begin{equation}
\vec{p}_a + \vec{p}_b - {\tilde L}_p (E_a \vec{p}_b + E_b \vec{p}_a)/c
- \vec{p}_c -\vec{p}_d + {\tilde L}_p (E_c \vec{p}_d + E_d \vec{p}_c)/c
\simeq 0
~.
\label{amel-conservnewp}
\end{equation}
It is easy to verify using the DSR1 boost
generators that when satisfied in one inertial frame
these conservation rules are also satisfied in all other
inertial frames.
It is noteworthy that these conservation laws ``mix" the particles (the
nonlinear correction terms involve properties of pairs of particles).
Other consistent candidates mixing the particles
and some consistent examples of non-mixing
laws are discussed in Refs.~\cite{amel-dsr1,polon2pap}.

\noindent
{\bf (DSR1.j): space-time sector.}
The DSR1 postulates concern energy-momentum space, but of course
one would like to identify a consistent space-time counterpart.
This identification process must still be considered as ``in progress";
however, various arguments~\cite{amel-dsr1,polon2pap},
and particularly the mentioned one-particle-sector
energy-momentum-space
connection with $\kappa$-Poincar\`{e} Hopf algebras,
suggest
that the $\kappa$-Minkowski noncommutative
spacetime, $[x_j,x_k]=0~,~~[x_j,t]=i {\tilde L}_p x_j$,
would provide a spacetime realization of DSR1.
To provide motivation for this identification we
just mention here that it is known~\cite{gacmaj}
that propagation in $\kappa$-Minkowski satisfies
the DSR1 dispersion relation and that products
of ``time-to-the-right-ordered wave exponentials"
in $\kappa$-Minkowski satisfy the property
$\left(e^{i \vec{p} {\cdot} \vec{x}} e^{-i E t} \right)
\left(e^{i \vec{k} {\cdot} \vec{x}} e^{-i \omega t} \right)=
\left(e^{i \vec{q} {\cdot} \vec{x}} e^{-i (E+\omega) t} \right)$,
with $\vec{q} = \vec{p} + e^{{\tilde L}_p E} \vec{k}
\simeq \vec{p} + \vec{k} + {\tilde L}_p E \vec{k}$.
This shows that the way to compose $\vec{p}$ and $\vec{k}$
into $\vec{q}$ resembles the structure of Eq.~(\ref{amel-conservnewp})
(but the correspondence is not exact and this requires
further study~\cite{polon2pap}).

\noindent
{\bf (DSR1.k): testability.}
DSR1 is definitely testable with forthcoming experiments
such as AMS~\cite{ams} and GLAST~\cite{glast},
which, as mentioned in Section~2, will test
very accurately the possibility of a wavelength/energy dependence
of speed of photons.
In fact, as discussed in point (DSR1.g),
DSR1 predicts the type of ${{L}_p E}$ correction to the speed-of-light
law that these experiments can verify.

For the other class of sensitive Lorentz-invariance tests
mentioned in Section~2, studies of particle-production thresholds
through observations of cosmic rays,
it appears~\cite{amel-dsr1,amel-dsr2,polon2pap}
instead that the effects predicted by DSR1 are not large enough
for testing with planned experiments.
This conclusion appears to be inevitable~\cite{amel-dsr1,amel-dsr2,polon2pap}
if energy-momentum deformed-conservation laws,
{\it e.g.}
Eqs.~(\ref{amel-conservnewe}),(\ref{amel-conservnewp}),
 are applied to
the DSR1 description of photonpion production (relevant for
observations of cosmic rays).

\section{Main features of DSR2}

\noindent
{\bf (DSR2.a): general structure.}
As mentioned, a second example of DSR theory, which we here call DSR2,
was more recently proposed in Ref.~\cite{leedsr} by Magueijo and Smolin.
It appears that a key guiding intuition for this study was the one
that a better example of DSR theory could be obtained by
describing the deformed boost
generators as a combination of the conventional
special-relativistic boost generators and of the
generator of dilatations.

\noindent
{\bf (DSR2.b): generators of deformed Lorentz transformations.}
In DSR2 (just like in DSR1) boost generators are deformed but
there is no deformation
of the rotation generators.
The differential representation of the DSR2 boost
generators is~\cite{leedsr} (again, without loss of
generality, we choose to focus on the boost that acts
along the $z$ axis):
\begin{equation}
N_z \simeq  p_z
\frac{\partial}{\partial E}
+E \frac{\partial}{\partial p_z}
- \lambda p_z \left(E \frac{\partial}{\partial E}
+ p_j\frac{\partial}{\partial p_j} \right )
~.
\label{generatDSR2}
\end{equation}
For conceptual clarity we are using a different notation, $\lambda$,
for the DSR2
deformation scale,
while for the analogous DSR1 scale we used ${\tilde L}_p$
(also notice that our $\lambda$
is the scale denoted by $l_0$ in Ref.~\cite{leedsr}).

\noindent
{\bf (DSR2.b): deformed Lorentz transformation rules.}
The DSR2 transformation rules were obtained explicitly
in Ref.~\cite{leedsr},
using an elegant argument which allowed to avoid explicit integration
of differential equations.
In particular, it was shown that
the DSR2 boost transformation rules
for the simple illustrative case
of a transformation
which is purely a boost in the $z$ direction and acts on a four-momentum
with components $(0,0,p_{z,0},E_0)$
should take the form:
\begin{equation}
p_z(v)= {\gamma(v) [p_{z,0} - v E]
\over 1 + \lambda [\gamma(v)-1] E_0 - \lambda \gamma(v) v p_{z,0}}
~,
\label{transfsdsr2p}
\end{equation}
where $\gamma(v) \equiv 1/\sqrt{1-v^2}$.

The parameter $v$ was interpreted
in Ref.~\cite{leedsr} as the relative velocity between the observers
connected by the boost
(but this appears, as discussed below, to lead to a questionable result
concerning the relation
between velocity and energy of a particle for a given observer).

The DSR2 formula (\ref{transfsdsr2p}) appears to be
very different from the corresponding DSR1 formula (\ref{pxiDSR1}).
Moreover, the DSR2 deformation, codified in (\ref{generatDSR2}),
appears to be very different from the DSR1 deformation,
codified in (\ref{generatDSR1exact}).
As mentioned this has led to various
speculations~\cite{judesvisser,polonDSR2nogood,dsrNATURE,chakdsr,ukDSR2,belgiumDSR2}
on the possible differences there might be present in the conceptual
structure of the two relativistic theories.
However, as we were approaching the study here reported
it became soon obvious that a genuine comparison between DSR1 and DSR2
could not be based on naive inspection
of Eqs.(\ref{transfsdsr2p}) and (\ref{pxiDSR1}),
which assume a different parametrization,
and could not even be based on naive inspection
of Eqs.~(\ref{generatDSR2}) and (\ref{generatDSR1exact}),
since the form of the generators of boosts is affected
by various aspects of a DSR theory, some
elementary aspects and some more delicate ones.
[The generators of boosts
encode (in an appropriate sense) at least two aspects of the DSR theory:
the aspect here of interest (the dependence of
momentum on rapidity) and the dispersion relation,
which deserves being considered separately.]

In order to have more robust ground for the comparison,
it appeared necessary to analyze the transformation laws
predicted by the two theories from the same perspective.
We then observed that the DSR2 boost generators (\ref{generatDSR2})
lead to the first-order differential equations
(focusing again, for simplicity, on the case in which
the components of momentum orthogonal to the boost vanish)
\begin{equation}
\label{sist2}
\left\{
\begin{array}{l}
\frac{\partial p(\xi)}{\partial \xi}=p'(\xi)=E(\xi)-\lambda p^2(\xi)\\
\\
\frac{\partial E(\xi)}{\partial \xi}=E'(\xi)=p(\xi)-\lambda E(\xi)p(\xi)\\
\end{array}
\right.
\end{equation}
which specify the relation between energy and momentum and rapidity, {\it i.e.}
they are the analog of the DSR1 differential equations (\ref{sist1}).

Again, even at this stage of comparison, a naive inspection
of (\ref{sist1}) and (\ref{sist2}) would appear to indicate that
there can be very little in common between the DSR1 transformation
laws and the DSR2 transformation laws.
However, if now we combine the equations in
 (\ref{sist2}) into a single (second-order) differential equation
for $p(\xi)$ we obtain
\begin{equation}
p''+ \lambda^2 p^3 + 3  \lambda p p'-p = 0~.
\label{ptwo}
\end{equation}
which is identical
(upon the, already implicit,
identification of constants $\lambda \equiv {\tilde L}_p$)
to the Eq.~(\ref{p}) which we encountered at the corresponding
point of the analysis of DSR1.

Clearly DSR2, through a different path, turned out to reproduce
the same differential equation connecting momentum and rapidity
that was already at the basis of DSR1.
This striking fact should be further investigated. A key point
will be to establish how central is the role of the laws
governing the dependence of momentum on rapidity.
For some applications it is possible that the original linear
differential equations (separately of energy and momentum)
contain the relevant information. But in some cases, like in
the calculation of the dependence of momentum on rapidity,
one does not miss any information by combining the two
original linear
differential equations and using the dispersion relation to
obtain a single (second-order) differential equation.

\noindent
{\bf (DSR2.d): dispersion relation.}
The key point in which DSR1 and DSR2 differ is the dispersion relation.
Using (\ref{generatDSR2}) and the unmodified space-rotation generators
one easily verifies that
in DSR2 the deformed dispersion relation is codified in
\begin{equation}
m^2=\frac{E^2-|\vec{p}|^2}{(1-\lambda E)^2}
~,
\label{dispexactdsr2}
\end{equation}
which we also note here (for convenience in the
later study of the phenomenological
implications) approximated to leading order in $\lambda$:
\begin{equation}
m^2 = E^2 -\vec{p}^2 + 2 \lambda E (E^2-\vec{p}^2)
~.
\label{dispapproxdsr2}
\end{equation}

It is now clear that
DSR2 assigns the same relativistic law, $p(\xi)$, as DSR1, but adopts
a simple redefinition of the energy $E_{DSR2} = f(E_{DSR1})$,
\begin{equation}
E_{DSR2} =\frac{1-e^{-2\lambda E_{DSR1}} + \lambda^2 p^2}{2\lambda}
~,
\label{eredef}
\end{equation}
throught the analysis.
The DSR2 momentum depends on rapidity just as in DSR1 but
the redefinition of energy leads to a different dispersion relation
and a different dependence of energy on rapidity\footnote{But of
course the functions $E_{DSR1}(\xi)$ and $E_{DSR2}(\xi)$
differ in a very elementary way. Since the energy is constrained to
momentum by the dispersion relation, $E_{DSR1}(p)$, $E_{DSR2}(p)$,
and since momentum depends on rapidity in the same way in DSR1 and DSR2,
one basically finds $E_{DSR1}(\xi) = E_{DSR1}(p(\xi))$
and $E_{DSR2}(\xi) = E_{DSR2}(E_{DSR1}(p(\xi)))$, with $p(\xi)$ satisfying
(\ref{p}) in both cases.}.
It is easy to verify that one obtains a solution of the
DSR2 differential equations (\ref{sist2}) by simply assuming
that momentum depends on rapidity, $p(\xi)$,
as in the DSR1 case, {\it i.e.} according to (\ref{ptwo}),
and assuming that the dependence of energy on rapidity is also
specified by (\ref{ptwo}) through the DSR2 dispersion relation
(\ref{dispexactdsr2}) (basically obtaining $E(\xi)$ by composition
of the function $E(p)$, which is codified in the DSR2 dispersion
relation, with the function $p(\xi)$, which is codified in (\ref{ptwo})).

\noindent
{\bf (DSR2.e): mass.}
From the DSR2 dispersion relation it is easy to obtain the
relation between the ``casimir mass" $m$ and the rest energy $M$
\begin{equation}
m=\frac{M}{1-\lambda M}
~.
\label{massesdsr2}
\end{equation}
Notice that DSR2 (unlike DSR1) predicts that $M$
would differ from $m$ already in leading order:
\begin{equation}
m=M + \lambda M^2
~.
\label{massesdsr2approx}
\end{equation}

In Ref.~\cite{leedsr} it was argued that the physical mass of particles
should be identified in DSR theories with the casimir mass.
This concept of the ``physical mass" is troublesome.
We feel, adopting the viewpoint of Refs.~\cite{amel-dsr1,amel-dsr2},
that one could here contemplate 3
distinct concepts of physical mass: the ``rest-energy mass",
the inertial mass and the gravitational mass. For each of these 3 concepts
the physics community has introduced a dedicated operative procedure
giving physical meaning to the concept. In our present familiar theories
these 3 logically-independent concepts turn out to be identified.
This may or may not be true as we investigate new physical theories,
but we feel that at this point in the development of DSR theories
there is no reason to expect any anomaly in this respect: one can
straightforwardly identify the rest energy, and it is legitimate to
expect that the rest energy will still coincide with the inertial mass
and the gravitational mass.
The concept of ``casimir mass" is probably even troublesome at the
level of analysis in which one attributes operative meaning to mathematical
symbols: one would like to define the casimir mass roughly as the invariant
mass fixed by the dispersion relation, but actually this does not
single out the $m$ of Eq.~(\ref{dispexactdsr2}). $m$ is certainly
an invariant combination of energy and momentum, but also any
function of $m$, $m' \equiv f(m)$, is an invariant function of
energy and momentum. $M$, the rest energy, is a function of $m$ and
is itself an invariant function of
energy and momentum. It appears therefore legitimate to adopt $M$ also
as the ``casimir mass". But the key issue is that probably we shouldn't dwell
on the
casimir mass because it does not have any specific (operatively meaningful)
significance. We should focus on rest energy, inertial mass and gravitational
mass.
At this point only the concept of rest energy is clearly identifiable
in a DSR theory. And as long as we lack a full understanding
of the concept of inertial mass in DSR, it would be meaningless to
advocate deformations of the celebrated $E = m c^2$ Einstein relation,
which is most fundamentally a relation between the rest energy and
the inertial mass (which can be trivially extended, by boosting,
to a relation between the energy away from the rest frame and
the so-called ``relativistic mass").

\noindent
{\bf (DSR2.f): maximum momentum and energy.}
It is noteworthy that the structure of the DSR2 transformation rules
is such that, for positive $\lambda$, boosts saturate to a
maximum momentum $1/\lambda$ and correspondingly also energy
saturates to $1/\lambda$.
This is easily understood: momentum saturates just like in DSR1 because,
as it was exposed in our previous calculations, momentum behaves in DSR2
just like in DSR1; in addition, in DSR2 also energy saturates because
the simple map $E_{DSR1} \rightarrow E_{DSR2}$ that allows to obtain
DSR2 form DSR1 is a map such that $E_{DSR1} = \infty$ (which corresponds
to $p = 1/\lambda$) is mapped
into $E_{DSR2} = 1/\lambda$.
Since we now see that DSR1 and DSR2 share the same law $p(\xi)$
and it was known that both saturate momentum at the inverse of
the absolute length scale, it is natural to wonder whether
the combination of these facts would be inevitable:
If a DSR theory saturates momentum at the value $1/\lambda$
is it necessary that the theory is also characterized by
the specific law  $p(\xi)$ found in DSR1 (and DSR2)?
Our tentative answer is negative: by fixing the saturating behaviour
of a nonlinear function one is not able to specify the function.
However, it is possible that the logical consistency of a DSR
framework introduces (in a way not yet uncovered in the literature)
some additional constraints on the form of the relevant nonlinear
functions, and perhaps these constraints fully specify the
function $p(\xi)$ when a maximum value $p=1/\lambda$ is imposed.

\noindent
{\bf (DSR2.g): velocity.}
In Ref.~\cite{leedsr} a DSR2 relation between the velocity of a particle
and its energy-momentum was only discussed implicitly.
As mentioned in point (DSR2.e),
the boost parameter $v$ of the transformation
law (\ref{transfsdsr2p})
was interpreted in Ref.~\cite{leedsr} as the relative velocity between
observers. Then, later in Ref.~\cite{leedsr}, this velocity $v$
enters the relation between the energy-momentum of a particle for a
first observer that sees the particle at rest and the energy-momentum
of that same particle for a second observer moving with
velocity $v$ with respect to the first observer.
From that relation it follows straightforwardly
that Magueijo and Smolin are adopting the
formula $v_{particle} = p /E$ to relate the velocity of a particle
to its energy-momentum.

We suspect that this relation might turn out to be inconsistent
with the full structure (still to be developed) of the DSR2 theory.
In fact, it is easy to verify (using
the DSR2 dispersion relation)
that the velocity formula $v_{particle} = p /E$
is not consistent with the relation $v_{particle} = {dE / dp}$.
As already emphasized in point (DSR1.g),
we feel that at this stage of development of DSR theories
there appears to be no reason to assume that in DSR theories
the relation $v_{particle} = {dE / dp}$, which holds in
Galilei relativity and in Einstein's special relativity\footnote{In
Einstein's special relativity (but not in Galilei relativity)
it happens to
be true that ${dE / dp} = p /E$. Somehow the interpretation
of $v$ adopted in Ref.~\cite{leedsr} preserves the (derived)
relation $v_{particle} = p /E$, violating the (fundamental)
relation $v_{particle} = {dE / dp}$, which is deeply connected with
the concept of group veloocity for waves and the possibility of
a conventional Hamiltonian formulation of the theory.
The formal structure of DSR2 allowed to introduce a boost parameter $v$
playing a role with
some formal analogies with the role played by the relative velocity
between observers in Einstein's theory; however,
the intuitive (but unjustified)
assumption that the boost parameter $v$ of DSR2 should coincide
with the relative velocity between the observers connected
by the boost
leads to the peculiar result $v_{particle} \neq {dE / dp}$.}
(and, as mentioned, is crucial for the Hamiltonian
formulation for DSR theories),
should lose validity.

If our concerns on this aspect of the analysis reported
in Ref.~\cite{leedsr} are justified it would follow that
the boost parameter $v$ of the transformation
law (\ref{transfsdsr2p})
should not be interpreted as the relative velocity between
the observers connected by a DSR2 boost.
Consequently also the velocity law $v_{particle} = p /E$ should be incorrect
in DSR2.
We maintain, as done in Refs.~\cite{amel-dsr1,amel-dsr2},
that the most reasonable assumption at this stage of
development of DSR theories should be $v_{particle} = {dE / dp}$,
and notice that, according to this viewpoint,
the DSR2 velocity law is
\begin{equation}
v_{particle} = {dE \over dp} =
{\sqrt{E^2- (1-\lambda E)^2 m^2}
\over E -\lambda^2 m^2 E+ \lambda m^2}
 =
{ v \over 1 +
\lambda m \sqrt{1-v^2} }
=\frac{p-\lambda Ep}{E-\lambda p^2}
~,
\label{veloxdsr2}
\end{equation}
instead of $v_{particle} = p /E$.

For what concerns attempts to identify doable experimental
tests of DSR2 predictions it is important to notice that
in any case, both adopting $v_{particle} = {dE / dp}$ and
adopting $v_{particle} = p /E$, in DSR2 one finds that:

\noindent
$\bullet$ the speed of massless particles ({\it e.g.} photons)
is still $c$, as in Einstein's theory;

\noindent
$\bullet$ the speed of massive particles
is related to energy in DSR2 through a relation
that differs from the corresponding relation of Einstein's theory
only through corrections of magnitude $L_p m^2/E$ (generically
much smaller than the $L_p E$ velocity corrections of DSR1).

\noindent
{\bf (DSR2.h): magnitude of the deformation.}
As shown above, in DSR2
$m^2 \simeq E^2 -\vec{p}^2 + 2 \lambda E (E^2-\vec{p}^2)
\simeq E^2 -\vec{p}^2 + 2 \lambda E m^2$
{\it i.e.} the dimensionless quantity characterizing the strength of
the deformation of the dispersion relation is $\lambda m^2/E$.
However,  DSR2 does not appear to be describable
as a theory which in all its aspects deforms ordinary special relativity
at order $\lambda m^2/E$. In particular, the boost generators chosen
in Ref.~\cite{leedsr} deform the ordinary special-relativity
boost generators at order $\lambda E$ (but their order-$\lambda E$
correction terms balance each other in such a way that the
DSR2 deformed dispersion relation, which is only deformed at order
$\lambda m^2/E$, is left invariant).

So, at least at the formal level DSR2 is, like DSR1, a modification
of ordinary special relativity with leading-order corrections
of characteristic strength $\lambda E$.
However, as we show later in the paper,
while the structure of DSR1 does lead to effects that are observably
large for certain planned experiments, none of the predictions
of DSR2 appears to be testable in the foreseeable future,
and this is
partly due to the fact that the DSR2 dispersion relation
only has $\lambda m^2/E$ deformation terms.

\noindent
{\bf (DSR2.i): two-particle sector and conservation laws.}
All results on Lorentz transformations and relativistic invariance
reported in Ref.~\cite{leedsr} on the DSR2 scheme concerned the
one-particle sector.
As emphasized in Refs.~\cite{amel-dsr1,amel-dsr2,polon2pap}
(and here mentioned already in our point (DSR1.i))
the step from the one-particle sector to multi-particle sectors
is highly non-trivial in DSR theories (unlike ordinary special
relativity) because of the nonlinearities.

It is actually quite hard to give a full description even just
of the two-particle sector; however, as shown
in Refs.~\cite{amel-dsr1,amel-dsr2,polon2pap}
there is one aspect of the two-particle (and in general multi-particle)
sector which can be easily studied and leads to important insight
concerning the possible experimental tests for DSR theories.
One can in fact establish which are the types of energy-momentum
conservation laws that can be applied to collision processes,
particularly the ones involving particle production such as
the process $\gamma + \gamma \rightarrow e^+ + e^-$.
We consider here DSR2 conservation laws
in leading order\footnote{As emphasized already above,
in relation with experimental tests of DSR theories it is always
sufficient to analyze the theories in leading order in the
second observer-independent scale; in fact, the new effects are
extremely small and therefore we might only
have a chance (if any) to see the leading-order corrections
(absolutely impossible to gain experimental insight
on the higher order corrections).}
in the deformation scale $\lambda$.

We find that in DSR2 in leading order in  $\lambda$ acceptable
energy-momentum conservation rules
for a generic $a + b \rightarrow c + d$ are
\begin{equation}
E_a + E_b + \lambda E_a^2 + \lambda E_b^2 -
E_c - E_d - \lambda E_c^2 - \lambda E_d^2
\simeq 0
~,
\label{consEdsr2}
\end{equation}
\begin{equation}
\vec{p}_a + \vec{p}_b + \lambda E_a \vec{p}_a + \lambda E_b \vec{p}_b -
\vec{p}_c - \vec{p}_d - \lambda E_c \vec{p}_c - \lambda E_d \vec{p}_d
 \simeq 0
~.
\label{conspdsr2}
\end{equation}
In fact, it is easy to verify, using the DSR2 boost
generators, that when satisfied in one inertial frame
these conservation rules are also satisfied in all other
inertial frames.

Together with the ``no-particle-mixing" conservation laws
(\ref{consEdsr2}),(\ref{conspdsr2}),
we also found\footnote{The reasons for
the availability of alternative choices of conservation laws
in DSR theories have been discussed in Refs.~\cite{amel-dsr1,polon2pap}.}
(again to order $\lambda$)
the ``particle-mixing" conservation laws
\begin{equation}
E_a + E_b - 2 \lambda E_a E_b -
E_c - E_d + 2 \lambda E_c E_d
\simeq 0
~,
\label{consEdsr2mix}
\end{equation}
\begin{equation}
\vec{p}_a + \vec{p}_b - \lambda E_a \vec{p}_b - \lambda E_b \vec{p}_a -
\vec{p}_c - \vec{p}_d + \lambda E_c \vec{p}_d + \lambda E_d \vec{p}_c
\simeq 0
~,
\label{conspdsr2mix}
\end{equation}
which also have the property
that they are necessarily satisfied in all inertial frames
if satisfied in a given inertial frames.

\noindent
{\bf (DSR2.j): space-time sector.}
All results on DSR2 Lorentz transformations and relativistic invariance
reported in Ref.~\cite{leedsr} concerned
energy-momentum space.
A key issue for future DSR2 studies will be the identification
of a spacetime that is consistent with
the structure of the DSR2 energy-momentum space.
We conjecture that this spacetime realization of DSR2
will require the introduction of a quantum spacetime
supporting a nontrivial differential calculus.
If this conjecture is correct it would affect severely some of
the remarks on DSR2 field theories made in Ref.~\cite{leedsr}.

The fact that also the DSR2
one-particle-sector
energy-momentum space can be cast~\cite{jurekDSRexamples}
in a way that is consistent with the general structure
of $\kappa$-Poincar\`{e} Hopf algebras could suggest that
perhaps some formulation of $\kappa$-Minkowski noncommutative
spacetime would also provide a candidate noncommutative
spacetime for DSR2, in the same sense which we already
discussed earlier for DSR1.
This remains an open issue for future research. We only observe
that, while for DSR1 there is the connection with ``time-to-the-right-ordered
wave exponentials" mentioned earlier, in the literature
there has been so far no discussion of
a $\kappa$-Minkowski ordering prescription which would lead
to DSR2-type equations.

\noindent
{\bf (DSR2.k): testability.}
As mentioned above, the only two classes of observations
with some chance of seeing DSR effects are:
(a) velocity-law tests (most notably the ones based
on observations of gamma-ray bursts~\cite{amel-grbgac}),
and (b) threshold tests (in cosmic-ray physics).

Since DSR2 predicts no modification of the
velocity law for photons, clearly
the velocity-law tests planned by AMS and GLAST
would not be sensitive to DSR2 effects.
Even for massive particles the DSR2 velocity law
is only different from the ordinary one at order $\lambda m^2/E$,
well below the reach of any foreseeable related experiment
(if indeed $\lambda$ is of the order of the Planck length).

Having here derived (in point (DSR2.h))
conservation laws for collision processes for the DSR2
scheme, we are now also in a position to study the
predictions of DSR2 concerning particle-production
thresholds and verify whether DSR2 gives rise to
effects that would be observable in the physics
of cosmic rays.
This can be done in complete analogy with the discussion we
gave above for the corresponding analysis with DSR1,
and the reader can easily verify that the result is
again negative:
one finds a modification
of the photopion-production
threshold but the modification is not large enough
to be noticeable in cosmic-ray observations.

We conclude that neither velocity-law tests nor
threshold tests can reveal DSR2 effects (DSR2 effects are too small in those
contexts). It appears that none
of the predictions of DSR2 can be tested in the
foreseeable future.

\section{Aside on nonlinearity in DSR}\footnote{This Section,
which was not present in the first version of this manuscript
(http://arXiv.org/abs/hep-th/0201245v1), has been added
in order to address some of the issues raised
in Refs.~\cite{ahluOPERAT,grumi}.}
In our analysis clearly a key role was played by the fact that
both DSR1 and DSR2 are based, in the one-particle sector,
on nonlinear realizations of the Lorentz group.
The fact that the Lorentz group is still present, but is nonlinearly
realized in some DSR proposals, has led some authors~\cite{ahluOPERAT,grumi}
to argue that essentially the nonlinearity might be a formal artifact,
that one should unravel the nonlinearity by performing some nonlinear
redefinition of the energy-momentum variables, and that ``true" translations
would still have to satisfy linear commutation relations with boosts.

While our focus has been on the DSR1/DSR2 comparison, rather than
going back to considerations of the conceptual content of the DSR
framework (which has been discussed in detail in many publications),
we should stress here that some recent studies,
perhaps most notably the ones reported
in Refs.~\cite{jurekNONLINEARgood,alessfranc,kodadsr},
have exposed the fact that the arguments
advocated in Refs.~\cite{ahluOPERAT,grumi} clearly
assuming that the underlying spacetime be classical (or at least
commutative).
The studies
reported in Refs.~\cite{jurekNONLINEARgood,alessfranc}, elaborating on points
already explored in the early
references~\cite{amel-dsr1,amel-jurekdsr,amel-gacJR},
show that in certain noncommutative spacetimes
boosts must act nontrivially on the one-particle
sector (which is faithfully described by the algebra relations)
and on the multi-particle sector (which in these noncommutative spacetimes
is described through the coalgebra sector of a Hopf algebra of symmetries
of the type considered in
Refs.~\cite{amel-kpoinfirst,amel-majrue,amel-kpoinap}).
The relativistic-invariant length scale $\lambda$ (or ${\tilde L}_p$)
cannot be fully removed\footnote{At the cost of an awkward nonlinear
redefinition of energy-momentum~\cite{judesvisser}, one can remove $\lambda$
completely from the one-particle sector and could be misled into believing that
the scale $\lambda$ would have been ``redefined away". But instead
the same redefinition of energy-momentum that trivializes the one-particle
sector requires that the action of boosts on two-particle states,
described by the coalgebra sector~\cite{jurekNONLINEARgood,alessfranc},
involves nontrivially the scale $\lambda$.},
since this is the natural energy-momentum manifestation of the
fact that the commutation relations among spacetime coordinates,
characterized by the length
scale $\lambda$, are being introduced as a fundamental feature of spacetime.
For example in the $\kappa$-Minkowski noncommutative spacetime
a proper description of translations
necessarily requires some new structures, as
most elementarily seen~\cite{jurekNONLINEARgood,alessfranc} by looking
at the product of plane waves in $\kappa$-Minkowski:
$[e^{- i k x} e^{i E t}] {\cdot} [e^{- i q x} e^{i \omega t}]
=[e^{- i (k + e^{\lambda E} q) x} e^{i (E+\omega) t}]$.
Clearly the spacetime noncommutativity is leading to a law of composition
of energy
momentum $((k,E),(q,\omega)) \rightarrow ((k + e^{\lambda E} q),(E+\omega))$
which is not covariant under ordinary boosts and involves a clear nonlinearity.

And there starts to be growing evidence
that noncommutativity is not the only realization of the idea
of ``spacetime quantization" that leads to DSR-type implementation
of the Relativity Principle. In particular, in Ref.~\cite{kodadsr}
it was recently shown that in certain quantum-gravity
theories a quantum (Hopf-algebra), rather than classical, symmetry
is realized when the cosmological constant is nonzero. And
the  symmetries are still quantum rather than classical
in the limit of vanishing cosmological constant
(the limit in which these quantum symmetries, of
course, are Poincar\'{e}-like).
Again the nonlinearities of the boost action turn out
to be unavoidable: one appears to have some freedom to transfer
some of the nonlinearity from the one-particle to the two-particle
sector or viceversa, but at least some dependence on a new
relativistic-invariant length (momentum) scale is
unavoidable.

The DSR framework was introduced~\cite{amel-dsr1} as a contribution
to quantum-gravity research and it implicitly assumes
that spacetime has strikingly nonclassical features, such as noncommutativity.
By focusing, as essentially done in Refs.~\cite{ahluOPERAT,grumi},
on the inadequacy of the DSR framework
for theories in a classical spacetime one is basically missing the
key objective of the DSR proposal. DSR was never
intended to apply in a classical spacetime. In an ordinary classical
spacetime, in which
a familiar concept of ``translations" should be applicable, we
cannot imagine a good reason for the proper concept of energy-momentum
to require a nonlinear action of boosts on energy-momentum.
One could formally introduce by hand a nonlinearity (just by mapping the
standard energy-momentum into ``new" energy-momentum),
but this should not have any observable consequences since it
is simply a formal statement.
Instead, as we just illustrated above in considering the
product of plane waves in a noncommutative spacetimes,
in certain quantum spacetimes the concept of translations may
require profound revisions and the concept of action of boosts
on energy-momentum must also be revised accordingly.
The type of quatum spacetimes considered
in Refs.~\cite{jurekNONLINEARgood,alessfranc,kodadsr}
provide examples of natural application of the DSR framework.

\section{Summary and open issues}
DSR theories, theories of the relativistic transformation laws
between inertial frames which are characterized by two observer-independent
scales~\cite{amel-dsr1}, had been often discussed in terms of
three features: the
dependence of energy and momentum on rapidity, $E(\xi)$ and $p(\xi)$,
and the dispersion relation $E(p)$.
This is clearly redundant since
the functions $E(p)$ and $p(\xi)$
completely fix the function $E(\xi)$.
We have observed that the deformed boost generators of a DSR
theory actually codify simultaneously the two
functions $E(p)$ and $p(\xi)$. Two DSR theories can have very
different form of the boost generators, but in some cases this
formal difference might hide the fact that the two theories
share the same dispersion relation or the same dependence
of momentum on rapidity. We have shown that the relation
between DSR1 and DSR2,
the most studied examples of DSR theories, is actually
of the special type in which the two theories describe the
dependence momentum on rapidity through exactly the same differential
equation. This came as a surprise because the DSR2 boost generators
appear to be very different from the DSR1 boost generators. A posteriori,
we understand the difference between DSR2 boosts and DSR1 boosts
as due exclusively to a simple map between the concept of
energy adopted in the two theories $E_{DSR2}= f (E_{DSR1})$.
All results previously obtained for the DSR1 theory also apply
to DSR2, upon this simple redefinition of energy.

We now understand that the
fact that momentum saturates to $1/L_p$ in the infinite rapidity
limit in both theories is not accidental: the differential equation
that governs the dependence of momentum on rapidity is exactly
the same in DSR1 and DSR2.
The fact that in DSR1 energy diverges in the infinite-rapidity limit
(just like in ordinary special relativity) while in DSR2 energy
saturates
to $1/L_p$ in the infinite rapidity
limit is also easily understood: the map $E_{DSR2}= f (E_{DSR1})$
is such that an infinite value of $E_{DSR1}$ corresponds
to the value $1/L_p$ for $E_{DSR2}$. (For simplicity
of discourse, we are
here again implicitly assuming ${\tilde L}_p = \lambda = L_p$.)

Similarly, acceptable laws of deformed energy-momentum conservation
for DSR2 can be obtained, as here shown, by simply taking the
corresponding DSR1 laws and performing the straightforward
redefinition of energy.

The fact that DSR1 and DSR2 are connected by such a simple relation
naturally suggests further scrutiny
of the points made in previous studies
in which a sharp difference, even of conceptual nature,
was drawn between DSR1 and DSR2.
These two DSR theories are very similar, perhaps too similar to
be treated separately.
We have here provided a unified language of notations and conventions
which we hope to prove useful as these theories are investigated.

The fact that we uncovered the very simple relation
between DSR1 and DSR2 also put us in the position to question
the puzzling fact that some key concepts, such as
particle mass and particle velocity,
were being investigated from very different perspective
in some of the studies focusing on DSR1 and
in some of the studies focusing on DSR2.
We found that the differences originated
in unjustified physical interpretation of the different choices
of notations and conventions which had so far been used
in studies of DSR1 and DSR2.
In particular,
we argued that also in DSR2, just like in DSR1, the only
meaningful concept of physical mass that can be so far
discussed is the one of rest-energy mass, and that in neither
of the two theories there is at present any ground for
speculating about modifications of the
relation $E_{rest} = m_I c^2$ between rest energy
and inertial mass (or speculating that there should be
differences between the rest-energy mass, the inertial mass
and the gravitational mass).
And we argued that in DSR2, just like in DSR1, there is so far
no scientifically robust argument in support of the idea
that the relation $v_{particle} =dE/dp$
(which has been so far always valid, both
in Galilei relativity and in Einstein's special relativity)
should be abandoned.
We have clarified that a contrary expectation which had
been presented in the literature actually originated
from the assumption that a DSR2-boost parameter,
suggestively denoted by ``$v$", should coincide with the relative
velocity of the two observers connected by the boost.
The only support for this interpretation is a vague
analogy with a corresponding parameter in
the description of ordinary special-relativistic
boosts, but we have shown that there is no robust
argument that could support this identification
in the context of the DSR2 theory.

Although the connection between DSR1 and DSR2
is of elementary conceptual nature,
we have shown that the energy redefinition that specifies
the connection can be quantitatively important in
certain contexts, leading to significantly different predictions.
Both DSR1 and DSR2, as so far constructed, are unable to
introduce a significant modification of the photopion-production
threshold relevant in the analysis of cosmic rays,
but DSR1 does predict a modification of the speed-of-light law
that could be observed in forthcoming experimental
studies~\cite{amel-grbgac,glast,gacQM100},
whereas DSR2 does not predict a modification of the
speed-of-light law.
It appears that none of the characteristic features of DSR2
could be tested in the foreseeable future.

Among the key open issues for the research here of interest
central importance should be attributed to the need
of endowing these relativistic theories with a suitable spacetime
sector. DSR theories of the DSR1/DSR2 type are introduced
in the energy-momentum sector. This is of course acceptable, but
eventually they must be extended to an associated spacetime sector
(just like in textbook presentations of ordinary special relativity
one introduces the concepts in the spacetime sector, but then
derives an associated energy-momentum sector).
We have stressed that these ``DSR spacetimes" should be ``quantum"
in an appropriate sense.
In fact, while the DSR idea of
two relativistic-invariant scales
could have other realizations~\cite{amel-dsr1},
DSR1 and DSR2, like basically all DSR theories in the literature,
are based on a nonlinear realization of the Lorentz group on
energy-momentum space.
We emphasized  that a nonlinear realization of the Lorentz group
on energy-momentum space
should find support in the structure of the associated spacetime.
It should be a manifestation of the fact that the ordinary concept
of translations is inadequate.
In an ordinary classical spacetime, in which
a familiar concept of translations should be applicable, we
cannot imagine a good reason for the proper concept of energy-momentum
to require a nonlinear action of boosts on energy-momentum.
It might be formally introduced, but only to eventually discover
that it didn't carry any physical significance.
Instead, as discussed in Section~4,
in a genuinely ``quantum" spacetime, for example
in a noncommutative spacetime, the concept of translations
can naturally require some revisions. In some noncommutative spacetimes
the structure of plane waves and of the commutators of coordinates
introduces an unavoidable nonlinearity in the law of composition
of momenta, and such a nonlinear composition law cannot possibly
be described covariantly in terms of a linear realization of
the Lorentz group, while it may well be covariant in the DSR sense.
For DSR1 the connection
with ``time-to-the-right-ordered wave exponentials"
in $\kappa$-Minkowski noncommutative spacetimes (which was here
mentioned in Section~2) should be further explored.
For DSR1, as mentioned in Section~3,
we are at an even more preliminary status in the search
of a realization in a quantum spacetime.

Our results could also motivate further studies of the
concept of maximum momentum in DSR theories, the aspect
of DSR theories which is most appealing from a quantum-gravity
perspective~\cite{amel-dsr1,leedsr}.
Both in DSR1 and DSR2 momenta saturate to $1/L_p$.
This is achieved through the same differential equation
governing the dependence of momentum on rapidity.
Are there other ways to achieve this saturation consistently
with the logical structure of DSR theories?
Assuming momenta do saturate to $1/L_p$ and that we are in
the context of a DSR theory, is the differential equation (\ref{p})
a robust general feature?

\section*{Acknowledgments}
We are grateful for conversations with A.~Agostini and G.~Mandanici.
Some of these conversations were particularly stimulating
during the transition from the first version of this manuscript
(http://arXiv.org/abs/hep-th/0201245v1) to the present version.
(This transition lasted a considerable amount of time because of
the desire to include some of the results then in preparation
for Ref.~\cite{andreatesi}.)

\end{document}